\documentclass[prl,a4paper,twocolumn,superscriptaddress,longbibliography
]{revtex4-2}
\usepackage{amssymb}
\usepackage{amsmath}
\usepackage{amsfonts}
\usepackage{graphicx}
\usepackage{bm}
\usepackage{xcolor,colortbl}
\usepackage{multirow}
\usepackage{float}
\usepackage{comment}
\usepackage{ulem}
\usepackage{relsize}
\usepackage{cmap}
\usepackage{bbold}
\usepackage{physics}
\usepackage{setspace}

\usepackage[colorlinks,citecolor=blue,linkcolor=red,urlcolor=blue]
{hyperref}

\definecolor{myGray}{gray}{0.9}

\usepackage[final]{pdfpages}
\usepackage{pgffor}

\makeatletter
\AtBeginDocument{\let\LS@rot\@undefined}
\makeatother

\DeclareMathOperator{\im}{Im}

\begin{document}

 	\title{Narrowing of 
  the flexural phonon spectral line in stressed crystalline two-dimensional materials 
  }
		
\author{A. D. Kokovin}	

\affiliation{Moscow Institute for Physics and Technology, 141700 Moscow, Russia}

\affiliation{\hbox{L.~D.~Landau Institute for Theoretical Physics, acad. Semenova av. 1-a, 142432 Chernogolovka, Russia}}

\author{V. Yu. Kachorovskii}
\affiliation{A. F.~Ioffe Physico-Technical Institute, 194021 St.~Petersburg, Russia}  
		
\author{I. S. Burmistrov}	

\affiliation{\hbox{L.~D.~Landau Institute for Theoretical Physics, acad. Semenova av. 1-a, 142432 Chernogolovka, Russia}}

\affiliation{Laboratory for Condensed Matter Physics, National Research University Higher School of Economics, 101000 Moscow, Russia}

\date{\today} 
	
\begin{abstract}

We develop the microscopic theory for the attenuation of  out-of-plane phonons in  stressed flexible two-dimensional crystalline materials. We demonstrate that 
the presence of nonzero tension strongly reduces the relative magnitude of 
the attenuation and, consequently, results in parametrical narrowing of the phononspectral line. We predict the specific power-law dependence of the spectral-line width on temperature and tension. We speculate that suppression of the phonon attenuation by nonzero tension might be responsible for high quality factors of mechanical nanoresonators based on flexural two-dimensional materials. 
\end{abstract}

\maketitle

Following the discovery of graphene~\cite{Novoselov2004,Novoselov2005,Zhang2005} and other atomically  thin  materials \cite{Novoselov2012}, flexible two-dimensional (2D)  materials \cite{2Dmat} have become at the focus of intense theoretical and experimental research.  These materials, being two-dimensional crystals, possesses  richer elastic physics due to an existence of out-of-plane deformations. 
Interplay of flexural strain and strong nonlinearities result in distinctive elastic properties -- known as anomalous elasticity. This includes non-trivial scaling  of elastic modules with a system size, crumpling transition with increasing  temperature 
and disorder, nonlinear Hooke's law, 
negative Poisson ratios, etc. \cite{Nelson1987,Aronovitz1988,Paczuski1988,David1988,Aronovitz1989,Guitter1988,Guitter1989,Toner1989,Doussal1992,Morse:1992,Nelson_1991,Radzihovsky1991,Morse:1992b,Bensimon_1992,Radzihovsky1995,Radzihovsky1998}. Recently, the theory of the anomalous elasticity of flexural materials has been further developed in various directions \cite{Kats2014,Gornyi:2015a,Kats2016,Burmistrov2016,Gornyi2016,Kosmrlj2017,Doussal2018,Burmistrov2018a,Burmistrov2018b,Saykin2020,Saykin2020b,Coquand2020,Mauri2020,Mauri2021,LeDoussal2021,Shankar2021,Mauri2022,Metayer2022,Burmistrov2022,Parfenov2022}.

For quite a while now, graphene and other flexural 2D materials are explored as nanoelectromechanical systems with unexpectedly high quality factors \cite{Miao2014,Leeuwen2014} (see also Refs. \cite{Steeneken2021,Ferrari2023} for a review). The observed quality factor increases on decreasing temperature. 
High quality factors of such mechanical nanoresonators suggest inefficiency of extrinsic  microscopic sources for 
damping 
\cite{Seoanez2007}. However, effective nonlinearity of 
the out-of-plane deformations (due to coupling between in-plane and
out-of-plane phonons)
provides intrinsic mechanism limiting magnitude of  the quality factor. Microscopically, an intrinsic quality factor of a flexural 2D material is intimately related with the attenuation of flexural phonons due to their interaction mediated by in-plane phonons. 

The presence of attenuation in the phonon spectrum in graphene on a substrate has recently been directly demonstrated experimentally by the method of the high resolution electron energy loss spectroscopy \cite{Li2023}. In spite of clear request for the theory of phonon attenuation in flexible 2D materials, we are aware of a single theoretical work addressing the attenuation of flexural phonons in a free standing membrane but of dimensionality $D{=}4{-}\epsilon$.

\begin{figure}[b]
\centerline{\includegraphics[width=1\columnwidth]{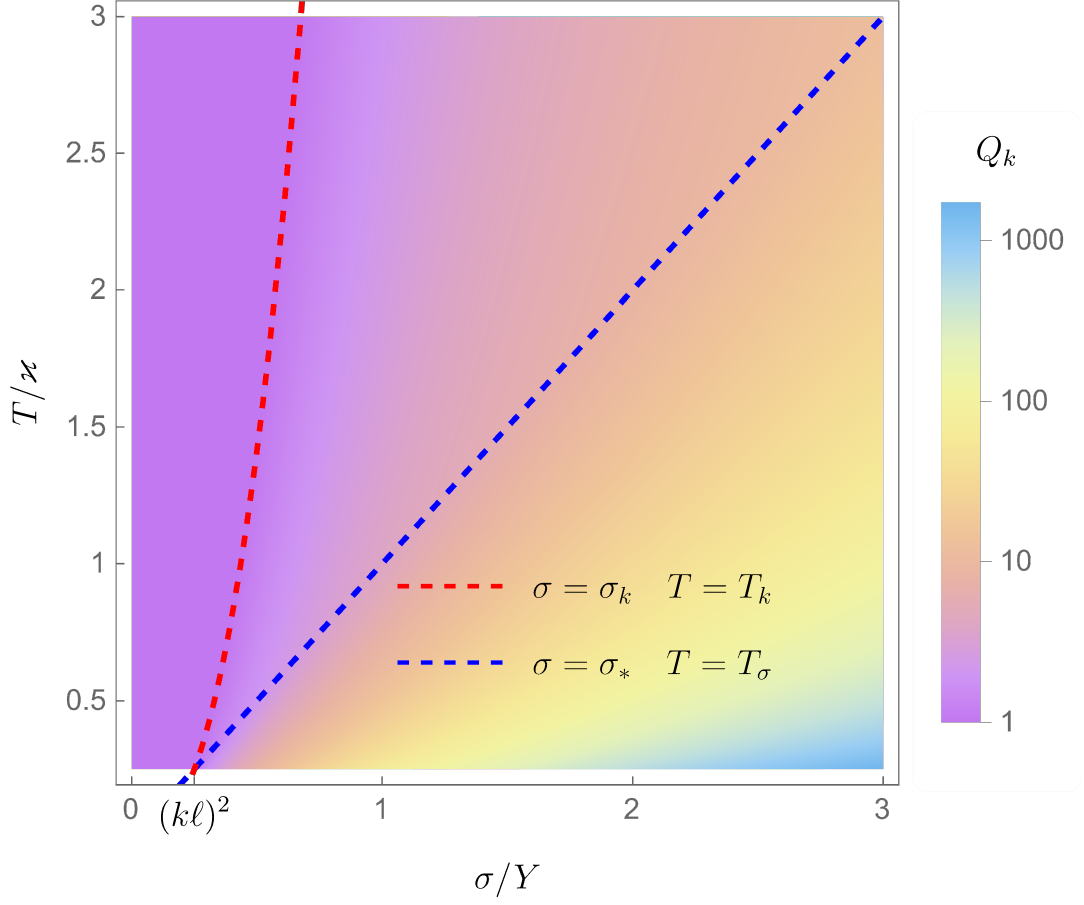}} 
\caption{The color density plot for dependence of the spectral line quality factor $Q_k {=}\tau_{k}\omega_{k}$ on 
temperature ($T$) and tension ($\sigma$), for $k\ell {\ll} 1$, 
 where $\ell{=}\sqrt{\varkappa/Y}$. Here we use the following temperature scales $T_k {\sim} \varkappa [(\sigma/Y)(k\ell)^{\eta{-}2}]^{2/\eta}$ and $T_\sigma{\sim}\sigma \varkappa/Y$. The corresponding scales for the tension are $\sigma_k{\sim}Y(k\ell)^{2{-}\eta}(T/\varkappa)^{\eta/2}$  and $\sigma_*{\sim}Y T/\varkappa$. (see text and Table \ref{tab:main}).}
\label{Figure:Fig:QF}
\end{figure}

Motivated by the experiment of Ref. \cite{Li2023}, in this Letter we develop microscopic theory for the attenuation of out-of-plane phonons in stressed flexural 2D materials. 
 Contact between a membrane and a substrate imposes a stress on 
 the membrane.
 Thus an existence of a nonzero tension acting on a 2D material is inevitable in
 a geometry of the experiment of Ref. \cite{Li2023}.
We focus on an experimentally relevant temperature range in which characteristics frequencies of flexural phonons are much smaller than the temperature, $T$.
We demonstrate that the presence of the tension strongly affects the flexural phonon attenuation and, consequently, the broadening of the phonon spectral line.
The point is that the tension enhances parametrically the real part of the spectrum of flexural phonons while not affecting the interaction between them. It is the latter that determines the phonon attenuation. Therefore, the ratio of the real and imaginary parts of the spectrum of flexural phonons enhances parametrically in the presence of the nonzero tension.  The predicted dependence of the spectral line quality factor on temperature, tension, and 
wave vector is shown in Fig. \ref{Figure:Fig:QF} (see also Table \ref{tab:main}). 
Also we discuss how our results for the phonon attenuation might be 
related with high quality factors 
observed in out-of-plane dynamics of a mechanical nanoresonator based on a flexural 2D material. We use units with $k_B{=}\hbar{=}1$ throughout the paper.

\noindent\textsf{\color{blue} Model.} --- The free energy describing thermal fluctuations in the flat phase of a 2D membrane can be written as 
$\mathcal{F} {=}\int d^2 \bm{x}\, [\varkappa  (\triangle\bm{r})^2/2{+}\sigma (\nabla \bm{r})^2/2]{+}\mathcal{F}_{\rm el}
$. Here $\varkappa$ is a bare  
bending rigidity while $\sigma$ is an external tension. We use a $d{=}d_c{+}2$ dimensional vector $\bm{r}$ to  parametrize a point on the membrane. The elastic crystalline energy, 
$\mathcal{F}_{\rm el}{=}\int d^2 \bm{x}\,[\mu \tr \hat{\bm{u}}^2 {+}\lambda (\tr \hat{\bm{u}})^2/2],$ is expressed via the $2{\times}2$ strain tensor $\hat{\bm{u}}$ with components 
$u_{\alpha\beta}{=}(\partial_\alpha \bm{r}\partial_\beta \bm{r}{-}\delta_{\alpha\beta})/2$. Here $\lambda$ and $\mu$ are the Lam\'e parameters of 2D crystalline material. Such elastic energy (with $d_c{=}1$) describes 2D crystalline materials with $D_{6h}$ (graphene) and $D_{3h}$ (transition metal dichalcogenide monolayers) point groups. Following \cite{David1988}, we will use a standard expansion in $1/d_c$ to efficiently describe anharmonic effects \cite{David1988}. To study phonon dynamics we employ the imaginary time action 
$\mathcal{S}{=}\int_0^\beta d\tau [\int d^2 \bm{x} \rho (\partial_\tau \bm{r})^2/2 {+} \mathcal{F}]$.
To exploit the flatness of the membrane, we use the parametrization of the coordinates with in-plane ($\bm{u}{=}\{u_x,u_y\}$) and out-of-plane ($\bm{h}{=}\{h_1,{\dots},h_{d_c}\}$) displacements: $r_1{=}\xi x {+} u_x$, $r_2{=}\xi y {+}u_y$, and $r_{a{+}2}{=}h_a$ with $a{=}1,\dots, d_c$. Here $0{<}\xi^2{<}1$ is the stretching factor which determines the ratio between projective and full area of the membrane. Following  \cite{Nelson1987},  we integrate out in-plane fluctuations $\bm{u}$ in harmonic approximation, and obtain the effective action for the flexural, out-of-plane phonons only:  
$\mathcal{S}_{\rm eff}{=}\mathcal{S}_\xi{+}\mathcal{S}_0{+}\mathcal{S}_{\rm int}$. Here 
$\mathcal{S}_0{=}\sum_{\omega k} [\rho \omega^2{+}\sigma k^2{+}\varkappa k^4 ]| h_{\bm{k},\omega}|^2/2$ describes non-interacting flexural phonons. The term \footnote{Here a `prime' sign in the last integral indicates that the interaction with $q{=}0$ is excluded. The so-called zero-mode contribution $\mathcal{S}_\xi{=}\int d^2\bm{x} \, c_{\alpha\beta} \varepsilon_\alpha\varepsilon_\beta/(8T)$, where $\varepsilon_\alpha{=}\xi^2{-}1{+}  \sum_{\omega,\bm{k}}
k^2_\alpha |h_{\bm{k},\omega}|^2$ and $c_{\alpha\beta}{=}\lambda {+}2\mu\delta_{\alpha\beta}$.}
\begin{equation}
\mathcal{S}_{\rm int} =\frac{Y}{8} 
\sum_{\Omega,\bm{q}{\neq} 0}
 \Biggl | \sum_{\omega,\bm{k}} \bm{k}_\perp^2 h_{\bm{k+q},\omega+\Omega}h_{-\bm{k},-\Omega}\Biggr |^2 .
\label{eq:action:2}
\end{equation}
is responsible for the interaction of the flexural modes mediated by the in-plane phonons.
The bare strength of interaction is determined by the Young's modulus $Y{=}4\mu (\mu {+} \lambda)/(2\mu {+} \lambda)$. Here we use the following notations  $\bm{k}_\perp{=}[\bm{k}{\times}\bm{q}]/q$ and $\sum_{\omega,k}{=}T\sum_\omega\int d^2\bm{k}/(2\pi)^2$.

\noindent\textsf{\color{blue} Anomalous elasticity.} The action $\mathcal{S}_0$ determines the bare spectrum of flexural phonons, $\omega_k{=}\sqrt{(\sigma k^2{+}\varkappa k^4)/\rho}$. In the statics it is well established that the perturbation theory in powers of RPA-type screened interaction (see below) produces ultra-violet logarithmic divergences that can be summed up by means of the renormalization group. Then, in the absence of the tension, $\sigma{=}0$, a power law renormalization of the bending rigidity and Young's modulus emerges at low momenta   \cite{David1988,Aronovitz1988},
\begin{equation}
\varkappa(k) {=}\varkappa (k L_*)^{-\eta},
\, Y(k) {=} Y (k L_*)^{-2{+}2\eta}, \quad k L_*\ll 1 ,
\label{eq:ren:12}
\end{equation}
Here $L_*{=}\sqrt{32\pi \varkappa^2/(3 d_c Y T)}$ is the so-called Ginzburg length.
The magnitude of the universal exponent $\eta{=}2/d_c{+}(73{-}68\zeta(3))/(27 d_c^2)+\dots$ is known for $d_c{\gg}1$ from analytics \cite{Saykin2020}. In the case of  $d_c{=}1$ numerics yields $\eta{\simeq}0.795{\pm}0.01$ \cite{Troster2013}. The static renormalization affects dynamical properties making the true spectrum of flexural phonons  
less soften, $\omega_k {\sim} (\varkappa/\rho)^{1/2}k^2 (k L_*)^{-\eta/2}$ \cite{Aronovitz1988}.
 
In a stressed membrane, $\sigma{>}0$, there exists an additional length scale $L_\sigma$. For $\sigma{<}\sigma_*{=}\varkappa L_*^{-2}$ that scale is given as   $L_\sigma{=}
L_* (\sigma_*/\sigma)^{1/(2{-}\eta)}$ while 
$L_\sigma{=}
L_* (\sigma_*/\sigma)^{1/2}$ for  $\sigma{\geqslant}\sigma_*$. Provided $\sigma{<}\sigma_*$, the power-law renormalizations \eqref{eq:ren:12} hold for the interval  $L_\sigma^{{-}1}{\ll} k{\ll} L_*^{{-}1}$ only. For region of small momenta, $kL_\sigma{\ll} 1$, the renormalization of the bending rigidity and Young's modulus become freezed to their magnitudes at $k{=}L_\sigma^{-1}$, \footnote{In this work we neglect weak logarithmic depedence of $\varkappa$ on momentum for $k L_\sigma{\ll}1$ \cite{Kosmrlj2016}.}
\begin{equation}
\varkappa_\sigma {=}\varkappa (L_\sigma/L_*)^{\eta},
\quad Y_\sigma{=} Y (L_\sigma/L_*)^{2{-}2\eta} .
\label{eq:ren:34}
\end{equation}
However, these static renormalizations of the bending rigidity and Young's modulus does not affect the spectrum of flexural phonons, $\omega_k{\simeq}\omega_k^{(\sigma)}{=}k\sqrt{\sigma/\rho}$, at low momenta,  $kL_\sigma{\ll}1$ \cite{Guitter1989,Aronovitz1989,Burmistrov2018a}. It is this regime of low momenta we will below focus on.

\begin{figure}[b]
\centerline{\includegraphics[width=0.45\textwidth]{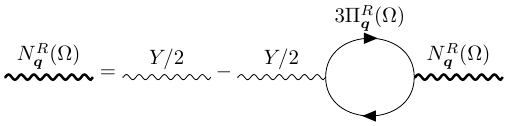}} 
\caption{The RPA-typ    e screened dynamical interaction. The wavy line denotes the bare interaction proportional to the Young's modulus. The solid line with arrow stands for the bare Green's function $G_{\bm{k}}(\omega)$.}
\label{Figure:Fig:RPA}
\end{figure}

\begin{figure}[t]
\centerline{\includegraphics[width=0.4\textwidth]{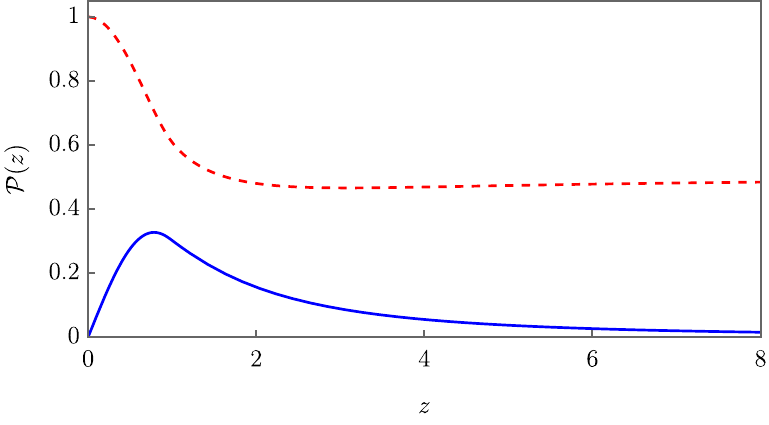}} 
\caption{The real (red dashed curve) and imaginary (blue solid) parts of $\mathcal{P}(z)$.}
\label{Figure:Pol}
\end{figure}

\noindent\textsf{\color{blue} Screened dynamical interaction.} Similar to the static case the strong bare interaction between flexural phonons has to be screened. The retarded RPA-type screened interaction becomes $N^{R}_{\bm{q}}(\Omega) {=} (Y_\sigma/2)/[1{+}3Y_\sigma\Pi^{R}_{\bm{q}}(\Omega)/2]$ instead of $Y_\sigma/2$ (see Fig. \ref{Figure:Fig:RPA}). We note that RPA is fully justified for $d_c{\gg}1$.  
The retarded polarization operator can be expressed in a standard way in terms of two bare Green's functions. At $q{\ll}L_\sigma^{-1}$ the polarization function acquires the scaling form
\begin{gather}
\Pi^{R}_{\bm{q}}(\Omega) {=} 
 \frac{2T d_c}{3} \int \!\frac{d\omega d^2\bm{k}}{(2\pi)^3}
\bm{k}_\perp^4 \Biggl \{G^{R}_{\bm{k}{+}\bm{q}}(\omega{+}\Omega)  \frac{\im G^{R}_{\bm{k}}(\omega)}{\omega}  \notag \\
{+}G^{A}_{\bm{k}}(\omega) \frac{\im G^{R}_{\bm{k}{+}\bm{q}}(\omega{+}\Omega)}{\omega{+}\Omega} 
\Biggr \} {\simeq} \frac{T  \mathcal{P}\left({\Omega}/{\omega_q^{(\sigma)}}\right)}{32\pi \varkappa_\sigma^2 L_\sigma^{-2}}
 .
 \label{eq:pol:R:0}
\end{gather}
Here we employed the classical limit of the boson equilibrium distribution function distribution 
since we focus on the high temperature regime, $T{\gg} \omega_q, |\Omega|$.
The bare retarded Green's function at $k{\lesssim} L_\sigma^{-1}$ is given as $G^R_{\bm{k}}(\omega){=}1/[\sigma k^2{+}\varkappa_\sigma k^4{-}\rho (\omega{+}i0)^2]$. For $z{\geqslant} 0$, the function $\mathcal{P}(z)$ is given by the integral representation (see Supplemental Materials \cite{SM})
\begin{equation}
\mathcal{P}(z) = \int_0^1 dx \, \mathcal{X}\bigl (z \sqrt{x}/(2-x)\bigr ) ,
\end{equation}
where
\begin{equation}
\mathcal{X}(z) = 1-2 z^2+\frac{4}{3} z^4+ \frac{4}{3} z |1-z^2|^{3/2} 
\begin{cases}
i , & |z|\leqslant 1\\
-1, & |z| >1  .
\end{cases}
\end{equation}
The real and imaginary part of the function $\mathcal{P}(z)$ are even and odd functions, respectively. They are shown in 
Fig.~\ref{Figure:Pol}. We note that $\mathcal{P}(z){\simeq} 1/2 {-}(2/(3z^2))[\ln z{-} i\pi/2]$ at $1{\ll}z{\ll}1/\sqrt{qL_\sigma}$. (For asymptotics at larger magnitudes of $z$ see \cite{SM}.)

In the limit of small momenta, $q L_\sigma{\ll}1$, and of not too large frequencies, $|\Omega|/\omega_q{\ll}\sqrt{\sigma_*/\sigma}$, the screening dominates, $Y_\sigma\Pi^{R}_{\bm{q}}(\Omega){\gg} 1$. Therefore, in that regime the screened interaction acquires the universal form determined by the inverse polarization operator 
$N^{R}_{\bm{q}}(\Omega) {\simeq}1/[3\Pi^{R}_{\bm{q}}(\Omega)]$. Since the polarization operator $\Pi^{R}_{\bm{q}}(\Omega){\sim}d_c$, the screened interaction becomes also weak, ${\sim}1/d_c$, with respect to the number of out-of-plane modes.

\begin{figure}[t]
\centerline{\includegraphics[width=0.4\textwidth]{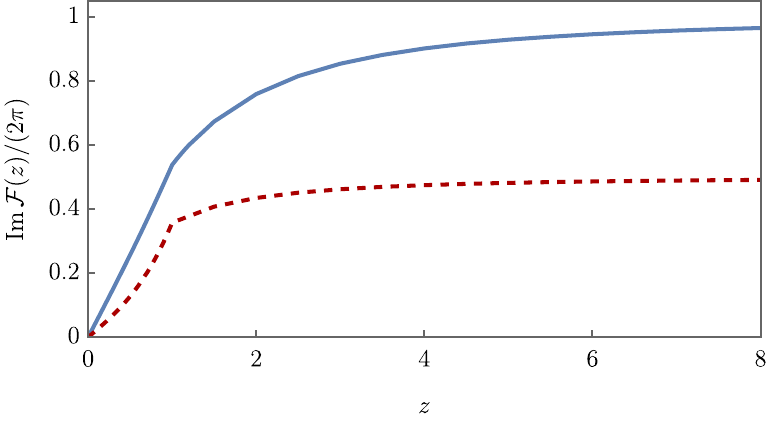}} 
\caption{The dependence of the imaginary part of the self-energy on frequency. Solid blue and dashed red curves are for the function $\mathcal{F}$ and $\tilde{\mathcal{F}}$, respectively.}
\label{Figure:Fig:ImF}
\end{figure}

\noindent\textsf{\color{blue} Flexural phonon attenuation at $\sigma{\ll}\sigma_*$.} The presence of imaginary part in $\Pi^{R}_{\bm{q}}(\Omega)$ results in the appearance of the imaginary part of the self energy for the exact Green's function of flexural phonons, $\mathcal{G}_{\bm{k}}^R(\omega){=}G^R_{\bm{k}}(\omega){-}\Sigma^{R}_{\bm{k}}(\omega)$. The latter is responsible for flexural phonon's attenuation. At $k{\ll}k_\sigma$, the lowest order self energy correction (see Fig. \ref{Figure:Fig:RPA})
acquires the scaling form, 
\begin{gather}
\Sigma^{R}_{\bm{k}}(\omega) {=} 
 {-}{4 T} \int \!\frac{d\Omega d^2\bm{q}}{(2\pi)^3}
\bm{k}_\perp^4 \Biggl[
G^R_{\bm{k}{+}\bm{q}}(\omega{+}\Omega) \frac{\Im N^R_{\bm{q}}(\Omega)}{\Omega} \notag \\ 
{+} 
N^A_{\bm{q}}(\Omega) \frac{\Im G^R_{\bm{k}{+}\bm{q}}(\omega{+}\Omega)}{\omega{+}\Omega} 
\Biggr ] {\simeq} \frac{\varkappa_\sigma k^4}{d_c} \mathcal{F}\left (\frac{\omega}{\omega_k}\right ).
\label{eq:Sigma:Pert:0}
\end{gather}
Since the self energy is proportional to $k^4$, its real part is completely negligible in comparison with the bare spectrum at $k L_\sigma{\ll}1$. The imaginary part of $\Sigma^R$, albeit being small, determines the attenuation of the flexural phonons. For the imaginary part of the function $\mathcal{F}(z)$, we obtain 
\begin{equation}
\Im \mathcal{F}(z) {=} \frac{8z}{3\pi}
\int \frac{d^2\bm{r}}{r^5} \frac{[\bm{n}{\times}\bm{r}]^4}{|\bm{n}{+}\bm{r}|^2}
\sum_{s=\pm} \Phi\left (\frac{z s{+}|\bm{n}{+}\bm{r}|}{r}\right ),
\label{eq:eq:F}
\end{equation}
where $\Phi(x){=}\Im\mathcal{P}(x)/[x |\mathcal{P}(x)|^2]$ and $\bm{n}$ is an auxiliary 2D unit vector. We find that $\Im \mathcal{F}(z){\simeq} 3.2
 z$ for $z{\ll}1$ and $\Im \mathcal{F}(z){\simeq}2\pi$ at $z{\gg} 1$. We note that Eq. \eqref{eq:eq:F} is applicable for $z{\ll}\sqrt{\sigma_*/\sigma}$ only since for larger magnitudes of $z$ one cannot use the universal form of the screened interaction (see \cite{SM}). The overall behavior of the imaginary part of the function $\mathcal{F}(z)$ is shown in Fig. \ref{Figure:Fig:ImF}. Since the real part of the self energy is proportional to $k^4$ it is completely negligible in comparison with $\sigma k^2$.

Next, defining the attenuation coefficient as $\gamma_{\bm{k}} {=} \Im\Sigma^R(\omega_{\bm{k}})/(\rho\omega_{\bm{k}}^2)$, we find 
$\gamma_{\bm{k}}{\simeq} (3.2/d_c) (k L_\sigma)^2{\ll}1$.  We note that this result holds at not too small $\sigma{\gg}\sigma_k{=}\sigma_*(kL_*)^{2{-}\eta}$. Indeed $L_\sigma$ becomes larger than
$1/k$ for $\sigma{\to} 0$. In such a regime, the effect of tension becomes negligible and one finds $\gamma_k{\sim} 1$ \cite{Kokovin2023l}.

\begin{figure}[b]
\centerline{\includegraphics[width=0.4\textwidth]{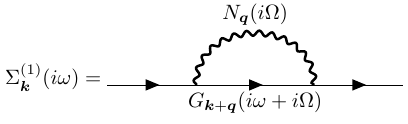}} 
\caption{The lowest order self energy correction.}
\label{Figure:Fig:Sigma}
\end{figure}

Although above we presented the result for $\Im \Sigma^R_{\bm{k}}(\omega)$ in the lowest order in $1/d_c$, they are in fact completely general. The point is that the exact Green's function satisfies the Ward-Takahashi identity, $\lim\limits_{\omega,k\to 0} [\mathcal{G}^{R}_{\bm{k}}(\omega)]^{-1} {=} \sigma k^2$ \cite{Guitter1989,Aronovitz1989,Burmistrov2018a}. It forbids corrections to the self-energy of the type $k^2 f(\omega/\omega_k)$. Therefore, the functional form for the self-energy correction given by Eq. \eqref{eq:Sigma:Pert:0} is quite general and holds for all $d_c{\geqslant} 1$. 

\noindent\textsf{\color{blue} Flexural phonon attenuation at $\sigma{\gg}\sigma_*$.} In that case, there is no power law renormalizations of $\varkappa$ and $Y$ since $L_\sigma{<}L_*$. In addition, the screening of interaction is weak and one can employ the following approximation $\im N^{R}_{\bm{q}}(\Omega) {\simeq}{-}3 (Y/2)^2 \im \Pi^{R}_{\bm{q}}(\Omega)$. We note that in this regime Eq. \eqref{eq:Sigma:Pert:0} reduces essentially to the Fermi golden rule expression for decay of the flexural phonon into three ones. Using Eq. \eqref{eq:Sigma:Pert:0}, we find 
\begin{equation}
\Sigma^R_{\bm{k}}(\omega){=} \left (\frac{3}{32\pi}\right )^2\left (\frac{L_\sigma}{L_*}\right)^4 \frac{\varkappa k^4}{2d_c} \tilde{\mathcal{F}}\left (\frac{\omega}{\omega_k}\right), 
\end{equation}
where $\im \tilde{\mathcal{F}}(z)$ is given by Eq. \eqref{eq:eq:F} but with the function $\tilde{\Phi}(x){=}\Im\mathcal{P}(x)/x$ substituted for $\Phi(x)$. The function has linear asymptote at $z{\ll}1$, $\im \tilde{\mathcal{F}}(z){\simeq} 1.42 z$ and tends to the constant at $1{\ll}z{\ll}1/\sqrt{qL_\sigma}$, $\im \tilde{\mathcal{F}}(z){\simeq}\pi$ (see \cite{SM}). The full dependence of $\im \tilde{\mathcal{F}}$ on $z$ is shown in Fig. \ref{Figure:Fig:ImF}. The above results implies that for $\sigma{\gg}\sigma_*$, the attenuation coefficient can be estimated as $\gamma_k {\sim} (Y/\sigma)^3(T/\varkappa)^2(k \ell)^2$, where $\ell{=}\sqrt{\varkappa/Y}$.

\noindent\textsf{\color{blue} Spectral line quality factor.} It is convenient to introduce the spectral line quality factor $Q_k$ as inverse of the attenuation coefficient, $Q_k{=}1/\gamma_k$. The $Q_k$-factor characterizes the quality of the resonance in phonon spectral function at $\omega{=}\omega_k$. The results for the attenuation coefficient discussed above (see Table \ref{tab:main}) suggests the following behavior of $Q_k$ for physical crystalline membrane (with $d_c{=}1$), 
\begin{equation}
Q_k {\sim} \begin{cases}
1,  & \! \sigma{\ll} \sigma_k \,\textrm{or}\, T_k{\ll}T,\\
\displaystyle \frac{(\sigma/Y)^{1{+}\alpha} (\varkappa/T)^\alpha}{(k\ell)^{2}},   & \! \sigma_k {\ll} \sigma {\ll} \sigma_*  \,\textrm{or}\, T_\sigma{\ll}T{\ll}T_k,\\
(k\ell)^{{-}2} (\sigma/Y)^3(\varkappa/T)^2,   & \! \sigma_*{\ll} \sigma  \,\textrm{or}\, T{\ll}T_\sigma.
\end{cases}
\label{eq:Qk:eq:main}
\end{equation}
Here 
we introduced two temperature scales $T_\sigma{=}\varkappa \sigma/Y$ and $T_k {\sim}\varkappa [(\sigma/Y)(k\ell)^{\eta{-}2}]^{2/\eta}$ corresponding to $\sigma_*$ and $\sigma$, respectively.

The temperature and tension dependences of $Q_k$ are controlled by the universal exponent $\alpha{=}\eta/(2{-}\eta){\simeq}0.67$. The overall behavior of the $Q_k$-factor with respect to $\sigma$ and $T$ is  
sketched in Fig. \ref{Figure:Fig:QF}. 
As naturally expected, the $Q_k$-factor increases with decrease of temperature. Interestingly, a nonzero tension induces not only dependence on the tension but also significant temperature dependence of $Q_k$. Figure \ref{Figure:Fig:QF} demonstrates clearly that $Q_k$ increases with increase of $\sigma$. Thus
the tension can serve as a tool to sharpen the phonon spectral line.

\noindent\textsf{\color{blue} Discussions.} Let us now estimate the spectral line quality factor $Q_k$. The characteristic tension $\sigma_*$ for graphene is equal approximately $0.1$ N/m. Taking build-in tension of a membrane to be $\sigma{=}10^{{-}2}$ N/m (it corresponds to relative deformation ${\sim} 10^{{-}4}$), we find $L_\sigma{\simeq} 10$ nm. We choose the momentum $k{\simeq} 0.01$ nm$^{{-}1}$ which corresponds approximately to $5{\cdot}10^{{-}4}$ fraction of the distance between $\Gamma$ and $K$ points in graphene. We note that such a magnitude is on the lower boarder of current resolution of electron energy loss spectroscopy technique \cite{Zhu2015}. Then we obtain the estimate $Q_k{\sim}10^2$, that implies extremely narrow phonon spectral line.

It would be tempting to apply our results for explanation of high quality factors of graphene-based mechanical nanoresonators \cite{Steeneken2021,Ferrari2023}. For a stressed membrane of size $L$, one could naturally associate the resonance frequency of its oscillation as $\omega_{k{\sim}1/L}^{(\sigma)}$. Then the spectral line quality factor 
$Q_{k{\sim}1/L}$ determines the quality factor of the resonance. For $L_\sigma{\ll} L$, we obtain a large quality factor $Q_{k{\sim}1/L}{\sim}(L/L_\sigma)^2{\gg} 1$. In addition, that quality factor has temperature dependence, $Q_{k{\sim}1/L}{\sim}T^{-\alpha}$. It resembles the $1/T$ dependence of the quality factor reported in the experiments. However, the above na\"ive estimates have to be taken with grain of salt. 

\begin{table}[t]
    \centering
\caption{The results for the real ($\omega_k$) and imaginary ($\gamma_k\omega_k$) part of the flexural phonon spectrum in different regions of $T$ and $\sigma$ (see Fig. \protect\ref{Figure:Fig:QF}).
Here we use the following temperature scales $T_k {\sim} \varkappa [(\sigma/Y)(k\ell)^{\eta{-}2}]^{2/\eta}$ and $T_\sigma{\sim}\sigma \varkappa/Y$. The corresponding scales for the tension are $\sigma_k{\sim}Y(k\ell)^{2{-}\eta}(T/\varkappa)^{\eta/2}$  and $\sigma_*{\sim}Y T/\varkappa$. Also we use $\varkappa_k{=}\varkappa (kL_*)^{-\eta}$ and $\ell{=}\sqrt{\varkappa/Y}$.}
\setstretch{1.2}
    \begin{tabular}{c|c|c}
         &  $\omega_k$ & $\gamma_k\omega_k$
         \\ \hline
 \rowcolor{myGray} $\sigma{\ll}\sigma_k$ or $T_k{\ll}T$  &  $k^2\sqrt{\varkappa_k/\rho}$ &$(\varkappa_k/\rho)^{1/2}k^2$ \\
 \hline
 \begin{tabular}{c}
 $\sigma_k{\ll}\sigma{\ll}\sigma_*$ \\
 or \\
 $T_\sigma{\ll}T{\ll}T_k$
 \end{tabular}& $k\sqrt{\sigma/\rho}$ &  $(\sigma/\rho)^{1/2} k (k\ell)^2 (Y/\sigma)^{1+\alpha}(T/\varkappa)^\alpha$\\
 \hline
 \rowcolor{myGray}
 $\sigma_*{\ll}\sigma$ or $T{\ll}T_\sigma$ & $k\sqrt{\sigma/\rho}$&  $(\sigma/\rho)^{1/2} k(k\ell)^2(Y/\sigma)^3(T/\varkappa)^2$
    \end{tabular}
    \label{tab:main}
\end{table}

As well known \cite{Sivan1994,Blanter1996,Altshuler1997,Mirlin1997,Silvestrov1997,Silvestrov2001,Auerbach2011,Gornyi2016s,Gornyi2017s}, the computation of broadening and decay for discrete levels is a   
tricky business. The phonon attenuation studied in our paper can be thought to be described alike Fermi's golden rule for a decay of a flexural phonon with frequency $\omega_k$ into other three flexural phonons with the probability amplitude proportional to the screened interaction $N^{R}_{\bm{q}}(\Omega)$. The corresponding three-particle level spacing can be estimated as 
\begin{equation}
\frac{1}{\Delta_3} {=} \prod\limits_{j=1}^3\int \frac{L^2d^2\bm{k_j}}{(2\pi)^2}
\delta\left (\omega_k {-} \omega_{{k_1}}{-}\omega_{{k_2}}{-}\omega_{{k_3}}\right ) {=} \frac{\omega_k^5 L^6 \rho^3}{5!(2\pi \sigma)^3} .
\end{equation}
In order the decay to be efficient and Fermi's golden rule to be applicable, the three particle level spacing has to be much smaller than the phonon spectrum attenuation, i.e. $\Delta_3{\ll}\gamma_k\omega_k$ \cite{Altshuler1997}. In the case, $L_*,L_\sigma{\ll}L$, the latter inequality transforms into the following condition $kL/\pi{\gg}(L/L_\sigma)^{1/4}\max\{1,(L_*/L_\sigma)^{1/2}\}$.
Since $kL{=}2\pi\sqrt{n^2{+}m^2}$ (where integers $n,m{\geqslant}1$ in the case of zero boundary conditions for $h$),  the quality factor of the resonance with $n,m{\gg}(L/L_\sigma)^{1/4}\max\{1,(L_*/L_\sigma)^{1/2}\}$ can be indeed  estimated as $Q_{k{\sim}1/L}$. In contrast, the applicability of our approach is questionable for low-frequency modes with $n{\sim}m{\sim}1$. We note, however, that discreteness can only enhance the quality factor $Q_{k{\sim}1/L}$. Therefore, our present result can be considered as a lower bound for $Q_{k{\sim}1/L}$ for for low-frequency modes with $n{\sim}m{\sim}1$. The detailed study of that situation is beyond the scope of present work \cite{elsewhere}.  

\noindent\textsf{\color{blue} Summary.} To summarize, we developed the theory for the attenuation of out-of-plane phonons in stressed flexible two-dimensional crystalline membranes.  We found that the presence of nonzero tension strongly reduces the relative magnitude of 
the attenuation coefficient and, consequently, results in 
narrowing of the spectral line. We predicted the specific dependence of the flexural-phonon spectral-line width on temperature and 
tension. Such dependence can be used to benchmark our theory in experiments on phonon spectrum measurements by means of the high resolution electron energy loss spectroscopy. We proposed that suppression of phonon attenuation due to nonzero tension can be responsible for high magnitude of the quality factors of mechanical nanoresonators based on flexural two-dimensional materials. 

\noindent\textsf{\color{blue} Acknowledgements.} The authors are grateful to Ya. Blanter and I. Gornyi for useful discussions. The work was funded in part by the Russian Ministry of Science and Higher Educations and by the Basic Research Program of HSE.

\bibliography{biblio-elasticity-fs}
	


\section*{Supplementary material}

# Narrowing of the flexural phonon spectral line in stressed crystalline two-dimensional materials

A. D. Kokovin,[1,2] V. Yu. Kachorovskii,[3] and I. S. Burmistrov[2,4]

[1]*Moscow Institute for Physics and Technology, 141700 Moscow, Russia*
[2]*L. D. Landau Institute for Theoretical Physics, acad. Semenova av. 1-a, 142432 Chernogolovka, Russia*
[3]*A. F. Ioffe Physico-Technical Institute, 194021 St. Petersburg, Russia*
[4]*Laboratory for Condensed Matter Physics, National Research University Higher School of Economics, 101000 Moscow, Russia*

In this notes we present details of calculation of (i) the polarization operator and (ii) the self-energy.

## S.I. DERIVATION OF THE EXPRESSION FOR THE POLARIZATION OPERATOR

The imaginary part of the equation (4) in the high-temperature limit ($T \gg |\omega|$, $|\Omega|$) is given by:

$$\text{Im}\,\Pi_{\boldsymbol{q}}^R(\Omega) = \frac{2T\Omega d_c}{3} \int \frac{d\omega}{2\pi} \int \frac{d^2\boldsymbol{k}}{(2\pi)^2} \boldsymbol{k}_\perp^4 \frac{\text{Im}\,G_{\boldsymbol{k}}^R(\omega)}{\omega} \frac{\text{Im}\,G_{\boldsymbol{k}+\boldsymbol{q}}^R(\omega+\Omega)}{\omega+\Omega} \ . \tag{S1}$$

Imaginary part of the retarded Green's function is

$$\text{Im}\,G_{\boldsymbol{k}}^R(\omega) = \frac{\pi}{2\rho\omega} \sum_{s=\pm 1} \delta(\omega + s\omega_{\boldsymbol{k}}). \tag{S2}$$

We proceed by integrating over $\omega$:

$$\text{Im}\,\Pi_{\boldsymbol{q}}^R(\Omega) = \frac{\pi T\Omega d_c}{12} \int \frac{d^2\boldsymbol{k}}{(2\pi)^2} \frac{[\boldsymbol{k}\times\boldsymbol{q}]^4}{q^4} \frac{1}{\rho\omega_{\boldsymbol{k}}^2} \frac{1}{\rho\omega_{\boldsymbol{k}+\boldsymbol{q}}^2} \sum_{s=\pm 1} \Big[ \delta\big(s\Omega + \omega_{\boldsymbol{k}} - \omega_{\boldsymbol{k}+\boldsymbol{q}}\big) + \delta\big(s\Omega + \omega_{\boldsymbol{k}} + \omega_{\boldsymbol{k}+\boldsymbol{q}}\big) \Big]. \tag{S3}$$

We will be interested in the region $q \ll q_\sigma \equiv L_\sigma^{-1}$. Therefore, $\omega_{\boldsymbol{q}} \simeq \omega_{\boldsymbol{q}}^{(\sigma)} = q\sqrt{\sigma/\rho}$. We make integral dimensionless by introducing new variables $z = \omega/\omega_{\boldsymbol{q}}$, $\boldsymbol{r} = \boldsymbol{k}/q$, $\boldsymbol{n} = \boldsymbol{q}/q$ and a small parameter $\alpha = q/q_\sigma \ll 1$:

$$\text{Im}\,\Pi_{\boldsymbol{q}}^R(\Omega) = \frac{\pi q^6 Tz d_c}{12\rho^2\omega_{\boldsymbol{q}}^4(4\pi^2)} \int d^2\boldsymbol{r}\, r^4 \sin^4(\varphi) \frac{\omega_{\boldsymbol{q}}^2}{\omega_{\boldsymbol{k}}^2} \frac{\omega_{\boldsymbol{q}}^2}{\omega_{\boldsymbol{k}+\boldsymbol{q}}^2} \sum_{s=\pm 1} \Big[ \delta\big(sz + \frac{\omega_{\boldsymbol{k}}}{\omega_{\boldsymbol{q}}} - \frac{\omega_{\boldsymbol{k}+\boldsymbol{q}}}{\omega_{\boldsymbol{q}}}\big) + \delta\big(sz + \frac{\omega_{\boldsymbol{k}}}{\omega_{\boldsymbol{q}}} + \frac{\omega_{\boldsymbol{k}+\boldsymbol{q}}}{\omega_{\boldsymbol{q}}}\big) \Big]$$

$$= \frac{Td_c}{64\pi\varkappa_\sigma^2 q_\sigma^2} \Big[ \mathcal{P}^{(1)}(\Omega/\omega_{\boldsymbol{q}}^{(\sigma)}) + \mathcal{P}^{(2)}(\Omega/\omega_{\boldsymbol{q}}^{(\sigma)}) \Big]. \tag{S4}$$

Here $\mathcal{P}^{(1,2)}$ corresponds to the contribution from the first and second delta-functions, respectively.

Let us start by evaluating the integral over the first $\delta$-function. The integral in this case will be dominated by $r \sim 1/\alpha$ ($k \sim q_\sigma$). Therefore we can neglect the difference between $\omega_{\boldsymbol{k}}$ and $\omega_{\boldsymbol{k}+\boldsymbol{q}}$ everywhere, except in the delta-function. The argument of the $\delta$-function is given by:

$$z = \pm \cos(\varphi) \frac{1 + 2r^2\alpha^2}{\sqrt{1 + r^2\alpha^2}} \tag{S5}$$

It is convenient to put $z > 0$ and then continue it to be an odd function. Integrating over $\cos(\varphi)$ will produce the same result for $s = \pm 1$:

$$\text{Im}\,\mathcal{P}^{(1)}(z) = \frac{16\alpha^2}{3} \int_0^\infty \frac{r\,dr}{(1 + r^2\alpha^2)^2} z \frac{\sqrt{1 + r^2\alpha^2}}{1 + 2r^2\alpha^2} \text{Re}\left(1 - z^2 \frac{1 + r^2\alpha^2}{1 + 2r^2\alpha^2}\right)^{3/2}. \tag{S6}$$

We proceed by introducing new variable $x = 1/(1 + r^2\alpha^2)$ and, then, derive the final result:

$$\text{Im}\,\mathcal{P}^{(1)}(z) = \frac{8}{3} \int_0^1 dx\, \frac{z\sqrt{x}}{(2-x)} \text{Re}\left(1 - \frac{z^2 x}{(2-x)^2}\right)^{3/2}. \tag{S7}$$

To obtain real part of the polarization operator, we employ Kramers-Kronig relation. Thus we find

$$\operatorname{Re}\mathcal{P}^{(1)}(z) = \int_0^1 dx\left[1 - 4u^2(x) + \frac{8}{3}u^4(x) - \frac{8}{3}u(x)\operatorname{Re}\left(u^2(x) - 1\right)^{3/2}\right], \quad u(x) = \frac{z\sqrt{x}}{(2-x)}. \tag{S8}$$

The asymptotics of $\mathcal{P}^{(1)}(z)$ at $z \gg 1$ can be found analytically:

$$\mathcal{P}^{(1)}(z) \simeq -4\frac{\ln z}{3z^2} + i\frac{2\pi}{3z^2}. \tag{S9}$$

Let us estimate the contribution from the second $\delta$-function in Eq. (S3). For $z>0$ and $s=-1$ its argument is zero only if $z>1$. For $z \ll 1/\alpha$ both $\omega_{\boldsymbol{k}}$ and $\omega_{\boldsymbol{k}+\boldsymbol{q}}$ can be substituted with $\omega_{\boldsymbol{k}}^{(\sigma)}$ and $\omega_{\boldsymbol{k}+\boldsymbol{q}}^{(\sigma)}$, respectively. We introduce the value of $r_0$ that nullifies the argument of the $\delta$-function:

$$r_0 = \frac{z^2-1}{2(z+\cos(\varphi))}. \tag{S10}$$

We can now rewrite the expression for $\operatorname{Im}\mathcal{P}^{(2)}$ as

$$\operatorname{Im}\mathcal{P}^{(2)}(z) = \frac{4\alpha^2 z}{3}\int dr d\varphi \frac{r^5\sin^4(\varphi)}{r^2(z-r)^2}\frac{\delta\left(r - \frac{z^2-1}{2(z+\cos(\varphi))}\right)(z-r)}{z+\cos(\varphi)}. \tag{S11}$$

Here we used the fact that the derivative of the argument of the delta function is given by:

$$|(z - r - \sqrt{r^2 + 1 + 2r\cos(\varphi)})'|_{r=r_0} = \frac{z+\cos(\varphi)}{z-r_0}, \qquad z - r_0 = \frac{z^2 + 2z\cos(\varphi) + 1}{2(z+\cos(\varphi))}. \tag{S12}$$

Now we perform integration over $r$ and obtain (for $1 \leqslant z \ll 1/\alpha$)

$$\operatorname{Im}\mathcal{P}^{(2)}(z) = \frac{\alpha^2 z(z^2-1)^3}{3}\int d\varphi \frac{\sin^4(\varphi)}{(z+\cos(\varphi))^3(z^2 + 2z\cos(\varphi) + 1)} = \alpha^2\frac{\pi}{3}z(z^2-1)^3\left\{\frac{2}{z} + \frac{3-2z^2}{(z^2-1)^{3/2}}\right\}. \tag{S13}$$

Let us compare the result (S7) and (S13). As one can see, for $z \sim 1$ the magnitude $\operatorname{Im}\mathcal{P}^{(2)}(z) \sim \alpha^2\operatorname{Im}\mathcal{P}^{(2)}(z) \ll \operatorname{Im}\mathcal{P}^{(1)}(z)$. However, for $z \gg 1$, we obtain from Eq. (S13) that

$$\mathcal{P}_2^{(2)}(z) \simeq \frac{\pi\alpha^2 z^2}{4}. \tag{S14}$$

It suggests, see Eq. (S9), that for $1/\sqrt{\alpha} \ll z \ll 1/\alpha$ the second contribution to the polarization operator dominates over the first one $\operatorname{Im}\mathcal{P}^{(1)}(z) \ll \operatorname{Im}\mathcal{P}^{(2)}(z)$.

For $z \gg 1$ the function $\operatorname{Im}\mathcal{P}^{(2)}(z)$ can be evaluated exactly at $\alpha \ll 1$. In this case, the condition of nullifying argument of the $\delta$-function is given by:

$$\alpha z - 2\alpha r\sqrt{1 + r^2\alpha^2} = 0. \tag{S15}$$

We note that in the considered region we can neglect the difference between $\omega_{\boldsymbol{k}}$ and $\omega_{\boldsymbol{k}+\boldsymbol{q}}$. Then, the integral can be exactly evaluated:

$$\operatorname{Im}\mathcal{P}^{(2)}(z) = p_2(\alpha z), \qquad p_2(x) = \frac{\pi x|x|}{(1 + \sqrt{1 + x^2})^2\sqrt{1 + x^2}}. \tag{S16}$$

We note that for $\Omega \gg \omega_q$ the second contribution to the polarization operator is independent of $q$. As one can see from Eq. (S16), we have that $\operatorname{Im}\mathcal{P}^{(2)}(z) \gg \operatorname{Im}\mathcal{P}^{(1)}(z)$ for $z \gg 1/\sqrt{\alpha}$ indeed.

To obtain the real part of $\mathcal{P}^{(2)}(z)$ we implement Kramers-Kronig relation. Thus we obtain

$$\operatorname{Re}\mathcal{P}^{(2)}(z) = p_1(\alpha z), \qquad p_1(x) = \frac{1}{x^2\sqrt{1+x^2}}\left[(2+x^2)\ln\frac{\sqrt{1+x^2}+1}{\sqrt{1+x^2}-1} + 4\sqrt{1+x^2}\ln\frac{|x|}{2}\right]. \tag{S17}$$

We emphasize that the real part of $\mathcal{P}^{(2)}(z)$ is not small at $z \ll 1/\sqrt{\alpha}$. In particular, we find,

$$p_1(x) \simeq 1 - (4\ln(x/2) + 3)x^2/8, \qquad x \ll 1. \tag{S18}$$

Therefore, at $z \ll 1/\sqrt{\alpha}$ the result for the polarization operator can be written as

$$\Pi_{\boldsymbol{q}}^R(\Omega) \simeq \frac{T d_c}{32\pi \varkappa_\sigma^2 q_\sigma^2} \mathcal{P}(\Omega/\omega_{\boldsymbol{q}}^{(\sigma)}), \qquad \mathcal{P}(x) = \frac{1}{2}\mathcal{P}^{(1)}(x) + \frac{1}{2}\mathcal{P}^{(2)}(0). \tag{S19}$$

It coincides with Eqs. (4)-(5) in the main text. For $z \gg 1/\sqrt{\alpha}$ the polarization operator becomes

$$\Pi_{\boldsymbol{q}}^R(\Omega) \simeq \frac{T d_c}{64\pi \varkappa_\sigma^2 q_\sigma^2} \Big[ p_1(\Omega/\omega_\sigma) + i p_2(\Omega/\omega_\sigma) \Big], \tag{S20}$$

where $\omega_\sigma = q_\sigma \sqrt{\sigma/\rho}$.

## S.II. DERIVATION OF THE EXPRESSION FOR THE THE SELF-ENERGY

The imaginary part of the self-energy is given as

$$\operatorname{Im}\Sigma_{\boldsymbol{k}}^R(\omega) = -2T\omega \int \frac{d\Omega}{\pi} \int \frac{d^2\boldsymbol{q}}{(2\pi)^2} \frac{[\boldsymbol{k}\times\boldsymbol{q}]^4}{q^4} \operatorname{Im} N_{\boldsymbol{q}}^R(\Omega) \frac{\operatorname{Im} G_{\boldsymbol{k}+\boldsymbol{q}}^R(\omega+\Omega)}{\Omega(\omega+\Omega)}. \tag{S21}$$

In the case $q_\sigma \ll q_*$ we can use the approximation $N_{\boldsymbol{q}}^R(\Omega) \simeq (3\Pi_{\boldsymbol{q}}^R(\Omega))^{-1}$. Then integrating over $\Omega$, we obtain:

$$\operatorname{Im}\Sigma_{\boldsymbol{k}}^R(\omega) = \frac{T\omega}{3} \sum_{s=\pm 1} \int \frac{d^2\boldsymbol{q}}{(2\pi)^2} \frac{[\boldsymbol{k}\times\boldsymbol{q}]^4}{q^4} \frac{\operatorname{Im}\Pi_q^R(\omega+s\omega_{\boldsymbol{k}+\boldsymbol{q}})}{\left|\Pi_q^R(\omega+s\omega_{\boldsymbol{k}+\boldsymbol{q}})\right|^2} \frac{1}{(\omega+s\omega_{\boldsymbol{k}+\boldsymbol{q}})\rho\omega_{\boldsymbol{k}+\boldsymbol{q}}^2}. \tag{S22}$$

As we will see below, the momentum integral is dominated by $q \ll q_\sigma$ (that is correct for $z = \omega/\omega_k \ll q_\sigma/k$), therefore we can use linear dispersion for $\omega_{\boldsymbol{k}}$. We then substitute the expression for the polarization operator from Sec. S.I and obtain

$$\operatorname{Im}\Sigma_{\boldsymbol{k}}^R(\omega) = \frac{8}{3\pi d_c} \omega\varkappa_\sigma^2 q_\sigma^2 \sum_{s=\pm 1} \int d^2\boldsymbol{q} \frac{[\boldsymbol{k}\times\boldsymbol{q}]^4}{q^4} \frac{1}{\omega_{\boldsymbol{q}}\rho\omega_{\boldsymbol{k}+\boldsymbol{q}}^2} \Phi_s, \tag{S23}$$

where

$$\Phi_s = \frac{2}{z\frac{\omega_{\boldsymbol{k}}}{\omega_{\boldsymbol{q}}} + s\frac{\omega_{\boldsymbol{k}+\boldsymbol{q}}}{\omega_{\boldsymbol{q}}}} \frac{\operatorname{Im}[\mathcal{P}^{(1)}(z\frac{\omega_{\boldsymbol{k}}}{\omega_{\boldsymbol{q}}} + s\frac{\omega_{\boldsymbol{k}+\boldsymbol{q}}}{\omega_{\boldsymbol{q}}}) + \mathcal{P}^{(2)}(z\frac{\omega_{\boldsymbol{k}}}{\omega_{q_\sigma}} + s\frac{\omega_{\boldsymbol{k}+\boldsymbol{q}}}{\omega_{q_\sigma}})]}{\left|\mathcal{P}^{(1)}(z\frac{\omega_{\boldsymbol{k}}}{\omega_{\boldsymbol{q}}} + s\frac{\omega_{\boldsymbol{k}+\boldsymbol{q}}}{\omega_{\boldsymbol{q}}}) + \mathcal{P}^{(2)}(z\frac{\omega_{\boldsymbol{k}}}{\omega_{q_\sigma}} + s\frac{\omega_{\boldsymbol{k}+\boldsymbol{q}}}{\omega_{q_\sigma}})\right|^2}. \tag{S24}$$

Here we note that functions with the superscript (2) have a small parameter $\omega_{\boldsymbol{k}}/\omega_{q_\sigma} = k/q_\sigma \ll 1$ in the argument. Since we are interested only in the leading order of the expansion with respect to parameter $k/q_\sigma \ll 1$, we can simply set this argument to zero. If we introduce new variables $\boldsymbol{r} = \boldsymbol{q}/k$, $\boldsymbol{n} = \boldsymbol{k}/k$ and rewrite the integral with the latter in mind, then we obtain

$$\operatorname{Im}\Sigma_{\boldsymbol{k}}^R(\omega) = \frac{8}{3\pi d_c} z\varkappa_\sigma k^4 \sum_{s=\pm 1} \int d^2\boldsymbol{r} \frac{[\boldsymbol{n}\times\boldsymbol{r}]^4}{r^5|\boldsymbol{n}+\boldsymbol{r}|^2} \Phi\left(\frac{z+s|\boldsymbol{n}+\boldsymbol{r}|}{r}\right), \qquad \Phi(x) = \frac{1}{x}\frac{\operatorname{Im}\mathcal{P}(x)}{|\mathcal{P}(x)|^2}. \tag{S25}$$

Here $\mathcal{P}(x)$ is given by Eq. (S19). We note, that this result is correct for $z \ll \sqrt{\sigma_*/\sigma}$. It is because we neglected 1 in comparison with $Y\Pi_{\boldsymbol{q}}^R(\Omega)$ in the denominator of $N_{\boldsymbol{q}}^R(\Omega)$. This holds only for $z \ll q_*/q_\sigma$. We finally obtain

$$\operatorname{Im}\Sigma_{\boldsymbol{k}}^R(\omega) \simeq \frac{\varkappa_\sigma k^4}{d_c} \mathcal{F}(z), \quad \mathcal{F}(z) = \frac{8}{3\pi} z \sum_{s=\pm 1} \int d^2\boldsymbol{r} \frac{[\boldsymbol{n}\times\boldsymbol{r}]^4}{r^5|\boldsymbol{n}+\boldsymbol{r}|^2} \Phi\left(\frac{z+s|\boldsymbol{n}+\boldsymbol{r}|}{r}\right). \tag{S26}$$

To obtain asymptotics for large $z$, we note, that integral is dominated by $r \sim z$. Therefore, we can approximately write:

$$\mathcal{F}(z) \simeq 2z \int_0^\infty \frac{dr}{r^2} \left\{ \Phi\left(\frac{z}{r} + 1\right) + \Phi\left(\frac{z}{r} - 1\right) \right\} = \int_{-\infty}^\infty dx \Big[ \Phi\left(x+1\right) + \Phi\left(x-1\right) \Big]$$

$$= \int_{-\infty}^\infty dx \left\{ \frac{1}{x+1}\operatorname{Im}\frac{1}{\mathcal{P}(x+1)} + \frac{1}{x-1}\operatorname{Im}\frac{1}{\mathcal{P}(x-1)} \right\} = \frac{2\pi}{\mathcal{P}(0)} = 2\pi. \tag{S27}$$

Here we used Eq. (S25) and Kramers-Kronig relation.

In case $q_\sigma \gg q_*$ the effect of screening is negligible. Therefore, we can substitute $\operatorname{Im} N_{\boldsymbol{q}}^R(\Omega) \simeq -3(Y/2)^2 \operatorname{Im} \Pi_{\boldsymbol{q}}^R(\Omega)$. Performing transformations as above, we obtain

$$\operatorname{Im} \Sigma_{\boldsymbol{k}}^R(\omega) \simeq \left(\frac{3q_*^2}{32\pi q_\sigma^2}\right)^2 \frac{\varkappa_\sigma k^4}{2d_c} \tilde{\mathcal{F}}(z), \quad \tilde{\mathcal{F}}(z) = \frac{8z}{3\pi} \sum_{s=\pm 1} \int d^2\boldsymbol{r} \frac{[\boldsymbol{n} \times \boldsymbol{r}]^4}{r^5 |\boldsymbol{n}+\boldsymbol{r}|^2} \tilde{\Phi}\left(\frac{z+s|\boldsymbol{n}+\boldsymbol{r}|}{r}\right), \quad \tilde{\Phi}(z) = \operatorname{Im} \mathcal{P}(z)/z.$$
(S28)

Asymptotic behaviour of $\tilde{F}(z)$ is similar to one of $F(z)$:

$$\tilde{F}(z) \simeq \pi.$$
(S29)

It holds for $z \ll 1/\sqrt{\alpha}$. The far asymptotics, $(z \gg 1/\alpha)$ is determined by the contribution from the second part of the polarization operator, see Eq. (S20). Then we obtain, that the correction to $\tilde{F}(z)$ for $z \gg \alpha^{-1}$ becomes

$$\delta\tilde{F}(z) = \frac{8}{3\pi} z \sum_{s=\pm 1} \int d^2\boldsymbol{r} \frac{[\boldsymbol{n} \times \boldsymbol{r}]^4}{r^4 |\boldsymbol{n}+\boldsymbol{r}|^2} \frac{\operatorname{Im} \mathcal{P}^{(2)}(z + s\sqrt{1+r^2+2r\cos(\varphi)})}{2(z+|\boldsymbol{n}+\boldsymbol{r}|)}.$$
(S30)

Assuming that typical $r \gg 1$, we find

$$\delta\tilde{F}(z) \simeq 2 \int_0^{\alpha z} \frac{dx \, p_2(x)}{x} + \int_{-\infty}^{\infty} \frac{dx \, p_2(x)}{|x - \alpha z|} = \pi \left(1 - \frac{2}{1+\sqrt{1+(\alpha z)^2}}\right) + \int_{-\infty}^{\infty} \frac{dx \, p_2(x)}{|x - \alpha z|}$$
(S31)

The last integral has logarithmic divergence at $x = \alpha z$. In the original variables it corresponds to the point $r = 0$ and comes from the substitution $|\boldsymbol{n}+\boldsymbol{r}|^2$ by $r^2$. It means that divergence of the denominator at $r \to 0$ stopped by $r \sim 1$. It can be explicitly seen from the following integral

$$\int_0^{2\pi} \frac{r d\varphi \sin^4\varphi}{1 + 2r\cos\varphi + r^2} = \frac{3\pi}{4} \begin{cases} 1/r - 1/(3r^3), & r \geqslant 1, \\ r - r^2/3, & 0 < r < 1. \end{cases}$$
(S32)

Let us now estimate the integral as (for $\alpha z \gg 1$)

$$\int_{-\infty}^{\infty} \frac{dx \, p_2(x)}{|x - \alpha z|} \simeq \int_0^{\infty} \frac{dx \, p_2(x + \alpha z)}{x} + \int_0^{\infty} \frac{dx \, p_2(\alpha z - x)}{x} \simeq \pi \int_\alpha^{\infty} \frac{dx}{(x + \alpha z)x} - \int_{-\alpha z}^{\infty} \frac{dx \, p_2(x)}{x + \alpha z}$$

$$\simeq \frac{\pi}{\alpha z} \ln\frac{\alpha z}{\alpha} - \int_{-\alpha z}^{\alpha z} \frac{dx \, p_2(x)}{x + \alpha z} - \int_{\alpha z}^{\infty} \frac{dx \, p_2(x)}{x + \alpha z} \simeq \frac{\pi}{\alpha z} \ln\frac{\alpha z}{\alpha} - \int_{-\alpha z}^{\alpha z} \frac{dx \, p_2(x)}{x + \alpha z} - \pi \int_{\alpha z}^{\infty} \frac{dx}{x(x + \alpha z)}$$

$$\simeq \frac{\pi}{\alpha z} \ln\frac{\alpha z}{\alpha} - \frac{\pi}{\alpha z} \ln\frac{2\alpha z}{\alpha z} - \int_0^1 \frac{du \, p_2(\alpha z u)}{u + 1} - \int_0^1 \frac{du \, p_2(\alpha z u)}{u - 1} \simeq \frac{\pi}{\alpha z} \ln\frac{z}{2} + \frac{2\pi}{\alpha z} \int_0^{1 - \alpha/(\alpha z)} \frac{du}{1 - u^2}$$

$$\simeq \frac{\pi}{\alpha z} \ln\frac{z}{2} + \frac{\pi}{\alpha z} \ln(2z) \simeq \frac{2\pi}{\alpha z} \ln z.$$
(S33)

Here we set lower limit for integration over $y$ by $\alpha$ since $y \sim \alpha r$. Therefore, for $\alpha z \gg 1$, we obtain with logarithmic accuracy

$$\delta\tilde{F}(z) \simeq \pi + \frac{2\pi}{\alpha z} \ln z.$$
(S34)

Therefore, at $z \gg 1/\alpha$ the function $\tilde{F}(z)$ tends to the value $2\pi$.



\end{document}